\documentclass[a4paper]{article}

\usepackage{19lomcon}        
\usepackage{cite}             
\usepackage{epsfig}           
\usepackage{epstopdf}
\begin{document}


\title{REVIEW OF STERILE NEUTRINO EXPERIMENTS}

\author{Seon-Hee Seo \email{sunny.seo@ibs.re.kr}
    }

\affiliation{Center for Underground Physics, Institute for Basic Science, 34126, Daejeon, Korea}


\date{January 2020}
\maketitle


\begin{abstract} 
There are $\sim3\sigma$ or more evidences of eV-scale sterile neutrinos from several different measurements. Many  dedicated experiments are (being) created and (will) take data to confirm or refute the eV-scale sterile neutrinos. In this talk, a mini review is presented on current experimental efforts and status on sterile neutrino search, especially using reactor and accelerator neutrinos. 
\end{abstract}

\section{Introduction}

The first experimental evidence of eV-scale sterile neutrino is from LSND (Liquid Scintillator Neutrino Detector) experiment in 2001 by observing 3.8$\sigma$ excess of $\overline{\nu}_{\rm{e}}$ events from a $\overline{\nu}_e$ appearance channel at 30~m baseline from 30~MeV $\overline{\nu}_{\mu}$ beam produced by muon decay at rest coming from 800~MeV proton beam on water target in Los Alamos National Lab. 

The main goal of MiniBooNE experiment at Fermilab was to confirm or refute the LSND result by observing $\nu_{\rm{e}}$ ($\overline{\nu}_{\rm{e}}$) appearance at 500~m baseline from 500~MeV $\nu_{\mu}$($\overline{\nu}_{\mu}$) beam produced by pion decay in flight coming from 8~GeV proton beam on Beryllium target. Note that MiniBooNE has a similar $L/E$ as LSND but different neutrino energy, signal, background and systematics. 
In 2009, MiniBooNE observed 3$\sigma$ excess of events in low ($<$475~MeV) energy region and this is consistent with the LSND excess but could not identify the excess is signal or unknown background. MicroBooNE was designed to identify the origin of the excess in MiniBooNE and currently has been taking data for few years.  

Another evidence of eV-scale sterile neutrino was observed in 2011 by radio-chemical solar neutrino experiments, GALLEX and SAGE, from 2.6$\sigma$ deficit of $\nu_{\rm{e}}$ events from $^{51}$Cr and $^{37}$Ar radioactive sources to calibrate their detectors.

Very short baseline (VSBL) reactor neutrino experiments in the 1990's also showed $\sim6\%$ deficit ($\sim$3$\sigma$) of $\overline{\nu}_{\rm{e}}$ events when compared to an updated calculation of reactor neutrino flux in 2011 by Mention et al~\cite{c:mention}. This deficit is so called RAA (Reactor Anti-neutrino Anomaly) and stimulated follow-up experiments even though there is a rather large uncertainty (2.4\%) of predicted reactor neutrino flux from Huber~\cite{c:huber} and Muller~\cite{c:mueller} model. 
The RAA best fit value with 3+1 $\nu$ scheme is $\Delta m^{2}_{41} = 2.4$~eV$^2$ and sin$^22\theta_{14} =$ 0.14. 
Recent VSBL reactor neutrino experiments are further motivated by the "5~MeV excess" observed by RENO, Daya Bay and Double Chooz in 2012 to identify the origin of the 5~MeV excess. 
In these VSBL experiments, background subtraction is done by subtracting reactor-OFF from reactor-ON data
assuming background is time invariant, and data analyses are performed in a model-independent way.

In the following sections, more details on current reactor and accelerator neutrinos experiments to search for an eV-scale sterile neutrinos are described. 
CEvNS experiments such as RED100, MINER, and CONUS etc, can also search for light sterile neutrinos as well as BEST, source experiment in BAksan, and KATRIN at Karlsruhe, but they are not discussed here due to limited space.

\section{Reactor $\nu$ Experiments for Sterile $\nu$ Search}

There are two types of detectors in VSBL reactor neutrino detectors. 
One is liquid scintillator (LS) and the other is plastic scintillator (PS) detector. 
LS detectors can be either homogeneous or segmented while PS detectors are all segmented. 
Segmented detectors are better to reduce background and to identify the vertex position. 
To further reduce background and to improve detection efficiency, 
gadolinium (Gd) or lithium ($^{6}$Li) are often loaded in one way or another in these detectors.

NEOS, Neutrino-4, STEREO, and PROSPECT are LS detectors and all are segmented and use research reactors except NEOS. 
DANSS and SoLid are PS detectors and use commercial and research reactors, respectively. 
NuLat and CHANDLER are mobile $\nu$ labs with PS detectors for nuclear monitoring at the surface and currently under R\&D and preparing a full detector, respectively.

Table~\ref{t:VSBL_exp} compares some important differences of these VSBL experiments. 
In the following subsections, more details on current VSBL experiments are discussed. 


\begin{table}[h]
  \centering
  \begin{tabular}{|l|r|r|c|c|c|}
  \hline
Experiment & Power & Baseline & Target mass & Target & Segment \\
           & [MW$_{th}$] & [m] & or volume & material & \\
\hline
NEOS       & 2800     & 24    & $\sim$1~m$^3$ & GdLS                  & No \\
Neutrino-4 & 100      & 6-12  & 1.8 ton  & GdLS                  & 2D \\
STEREO     & 57       & 9-11  & 2.4~m$^3$   & GdLS                  & 2D \\
PROSPECT   & 85       & 7-12  & 4 ton    & $^{6}\rm{LiLS}$         & 2D \\
DANSS      & 3000     & 10-12 & 1~m$^3$     & PS (Gd layer)         & 2D \\
SoLid      & 72       & 6-9   & 1.6 ton  & PS ($^{6}\rm{Li}$ layer)& 3D \\
NuLat*      & any & any   & 0.9 ton  & $^{6}\rm{LiPS}$         & 3D \\
CHANDLER*   & any & any   & $\sim$1 ton    & PS ($^{6}$Li layer)& 3D \\
\hline  
  \end{tabular}
  \caption{
Comparison of the current VSBL experiments.
  }
  \label{t:VSBL_exp}
\end{table}

\subsection{NEOS}

NEOS is a $\sim1m^3$ LS detector doped with 0.5\% Gd and located in a tendon gallery of the 5$^{\rm{th}}$ reactor (2.8~GW$_{\rm{th}}$) of the Hanbit nuclear power plant which is also used by RENO. The distance from the 5$^{\rm{th}}$ reactor core to the NEOS detector is 24~m and overburden is about 20 m.w.e. 
Total 38 8 inch PMTs are mounted on two side buffers and the energy resolution is measured to be about 5\% at 1~MeV. 

In phase-I, NEOS has taken 180 (46) days reactor-ON (-OFF) data from Aug. 2015 to May 2016 with $\sim$2000 IBDs/day where signal to background ratio is about 22 thanks to 10\% Ultima-Gold-F added to GdLS for good PSD (pulse shape discrimination).

The 5~MeV excess was clearly observed in this NEOS-I data set when compared with the Huber-Miller model. 
According to the shape analysis by comparing the prompt energy spectra of NEOS to Daya Bay, 
an oscillation pattern was seen but this was still within systematic uncertainty,
and therefore only an exclusion region was set. The RAA best fit was disfavored by NEOS-I at 4.3$\sigma$~\cite{c:neos}. 

NEOS phase-II has started taking data in September 2018 after refurbishing (new GdLS and muon veto system) its phase-I detector. Main goals of NEOS-II are ``rate and shape'' analysis for sterile neutrino search and spectral decomposition of $^{235}$U and $^{239}$Pu for a full fuel cycle of 500 days data, important for understanding origin of the 5~MeV excess.

\subsection{NEUTRINO-4}

Neutrino-4 is a segmented 1.8~m$^{3}$ LS detector doped with 0.1\% Gd and the GdLS is equally filled in optically isolated 10$\times$5 rectangular cells (W: 22.5~cm x H: 85~cm L: 22.5~cm) up to 70~cm height. 
Each of total 50 PMTs is mounted on the top of each cell and energy resolution is measured as 16\% at 1~MeV. 
Neutrino-4 detects neutrinos from SM-3 research reactor (100~MW$_{\rm{th}}$) in Russia with baseline from 6 to 12~m under a shallow overburden (3$\sim$5~m.w.e.). 

Neutrino-4 took 480 (278) days of reactor-ON (-OFF) data~\cite{c:nu4} where signal to background ratio is about 0.54 due to no PSD capability. Oscillation pattern in 3+1 $\nu$ scheme was observed with the best fit value of $\Delta m^{2}_{41} \approx 7$ eV$^2$ and sin$^22\theta_{14} \approx$ 0.4 at 3$\sigma$. However, this contradicts the absolute flux measurements at a longer baseline by Daya Bay and RENO due to much bigger deficit by Neutrino-4 at the averaged oscillation. 

Neutrino-4 plans to upgrade its detector to Neutrino-6 where new GdLS is filled with higher concentration of Gd and PSD capability. Expected sensitivity of Neutrino-6 is about 3 times better than current result of Neutrino-4. 

\subsection{STEREO} 

STEREO is a segmented 2.4~m$^{3}$ LS detector doped with 0.2\% Gd and the GdLS target is equally filled in 6 rectangular cells (W: 89.2cm, H: 91.8~cm, L:36.9~cm) optically isolated. The 6 cells of GdLS target are surrounded by two short LS cells in front and end and two long LS cells in both sides. 
Four 8 inch PMTs are mounted on top of each target cell and each short LS cell and energy resolution is measured as 9\% at 1~MeV. For each long LS cell, eight 8 inch PMTs are mounted on top of it. 
STEREO detects neutrinos from ILL research reactor (58~MW$_{\rm{th}}$) in France with baseline from 9 to 11~m under an overburden ($\sim$15~m.w.e.). 

STEREO took 179 (235) days of reactor-ON (-OFF) data with $\sim$400 IBDs/day where signal to background ratio is about 0.9. No excess around 5~MeV but deficit above 6~MeV was observed~\cite{c:stereo}. 
No oscillation pattern was also observed and the RAA best fit value was rejected at $>$ 99.9\%CL. 

\subsection{PROSPECT}

PROSPECT is a segmented 4~ton LS detector with 0.1\% $^6$Li doping and the $^6$LiLS is equally filled in 154 rectangular cells (14.5~cm x 14.5~cm x 117.6~cm per cell) optically isolated. 
A 5~inch PMT is attached to each side, apart by 117.6~cm length, of a cell and energy resolution is measured as 4.5\% at 1~MeV.
PROSPECT detects neutrinos from HFIR research reactor (80~MW$_{\rm{th}}$) in USA with baseline from 7 to 9~m under a very shallow overburden ($<$ 1~m.w.e.). 

PROSPECT took 33 (23) days of reactor-ON (-OFF) data with 771 IBDs/day where signal to background ratio is about 2.2 (1.32) for accidental (correlated) background thanks to good PSD power. 
No clear excess around 5~MeV was observed and no oscillation pattern was also observed due to large statistical uncertainty, and the RAA best fit value was rejected at 2.2$\sigma$~\cite{c:prospect}. 
PROSPECT is currently analyzing additional data (June-October 2018) with twice more IBD event statistics than the data used in current results.  

\subsection{DANSS}

DANSS is a segmented 1~m$^3$ PS detector consisting of 2,500 strips (W: 4~cm, H: 1~cm, L: 1~m per strip) of PS embedded with 3 wavelength shifting (WLS) fibers per strip coated with reflective material with 0.35\% Gd loading. 
Each of total 2,500 SiPMs is attached to one end of each strip, and additionally 50 PMTs are attached to 50 bundles of 50 strips resulting in $\sim$28\% energy resolution at 1~MeV. 
DANSS detects neutrinos from WWER-1000 reactor (3.1~GW$_{\rm{th}}$) in Russia with baseline from 10.7 to 12.7~m under an overburden of 50~m.w.e.

DANSS has been taking data since October 2016 with $\sim$5,000 IBDs/day (signal/background $\approx$ 33) at 10.7~m, and result with 193 days of reactor-ON data showed no excess around 5~MeV. 
When the prompt energy distribution at bottom position was compared to that of top position, small oscillation pattern was observed but did not match with the RAA best fit value which was rejected at 5$\sigma$~\cite{c:danss}. 
More recent preliminary result from DANSS~\cite{c:danss2} with about 2.5 times increased statistics showed some excess around 5~MeV region but still weak (1.8$\sigma$) evidence of 3+1 $\nu$ oscillation. 

\subsection{SoLid}

SoLid~\cite{c:solid} is a highly segmented 1.6~ton PS detector consisting of about 13,000 cubes (5x5x5~cm$^3$ per cube) of PS embedded with parallel and vertical optical fibers per cube covered with LiF:ZnS(Ag) sheets. 
Readout is done by total about 32,000 SiPMs operating at 5$^{O}$C and energy resolution is measured as $\sim$14\% at 1~MeV. 
It detects neutrinos from BR2 research reactor (50-80~MW$_{\rm{th}}$) in Belgium with baseline from 6 to 9~m under an overburden of 10~m.w.e. About 300 IBDs/day with $\sim$30\% efficiency is expected with a good signal to background ratio of three to one. Data has been taken for more than a year so far and the first physics result is expected to come out soon. 

\subsection{Daya Bay, RENO and JUNO}

\begin{figure}[h]
\begin{center}
\includegraphics[height=0.7\textwidth]{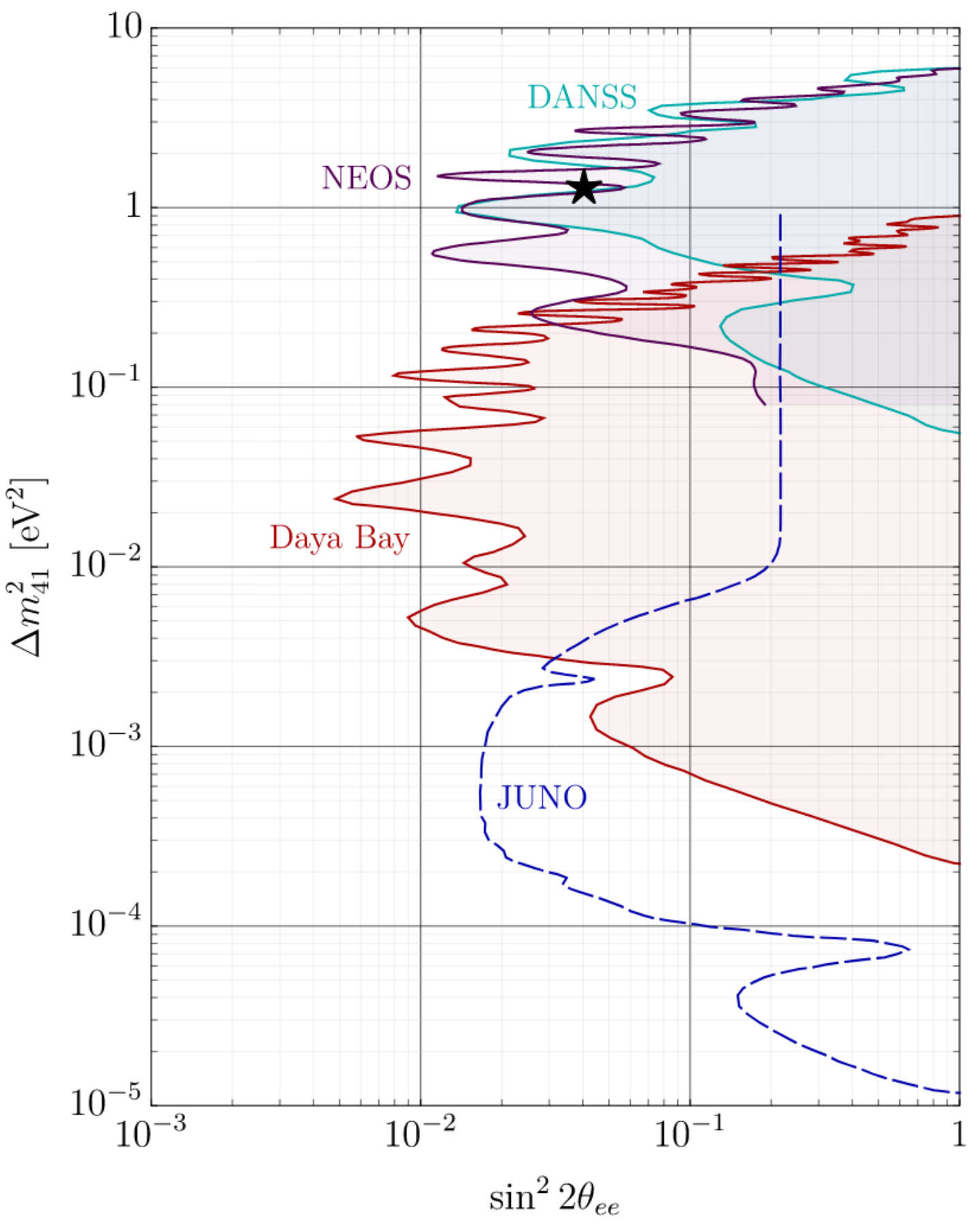}
\end{center}
\caption{
Exclusion limits and sensitivity for eV-scale sterile neutrino searches by NEOS, DANSS, Daya Bay and JUNO. Adapted from ~\cite{c:vsbl_sbl_limits}. 
}
\label{f:db_juno_sterileNu}
\end{figure}

Short baseline (SBL) reactor neutrino experiments, Daya Bay~\cite{c:db} and RENO~\cite{c:reno}, also searched for eV-scale sterile neutrinos assuming 3+1 $\nu$ scheme using 621 and 1,500 days data, respectively, with a Far-to-Near ratio method in their data analysis. Figure~\ref{f:db_juno_sterileNu}~\cite{c:vsbl_sbl_limits} shows NEOS, DANSS, and Daya Bay exclusion regions together with JUNO sensitivity. 
\section{Accelerator $\nu$ Experiments for Sterile $\nu$ Search}

\subsection{MiniBooNE}

\begin{figure}[h]
\begin{center}
\includegraphics[width=0.49\textwidth]{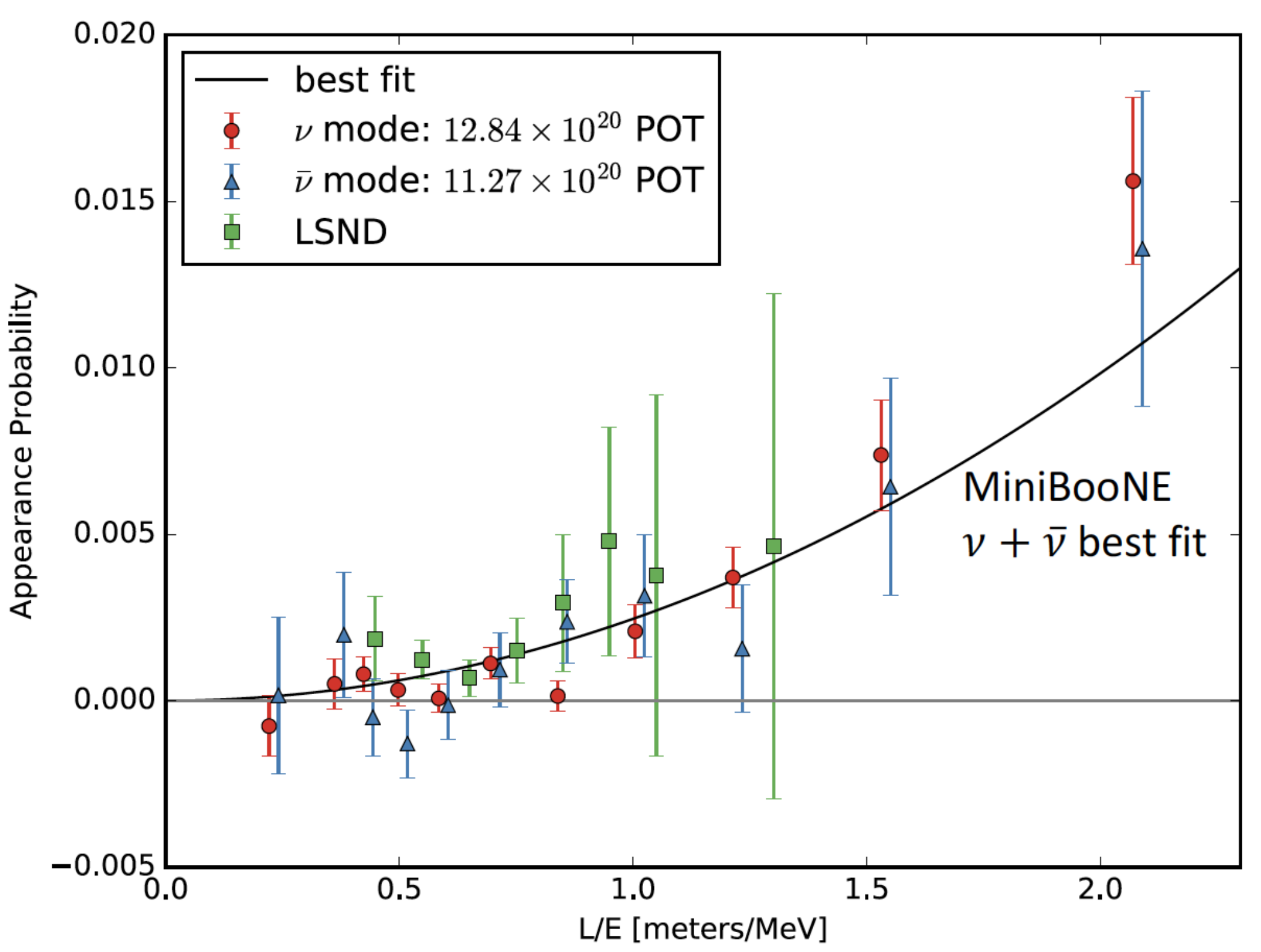}
\includegraphics[width=0.49\textwidth]{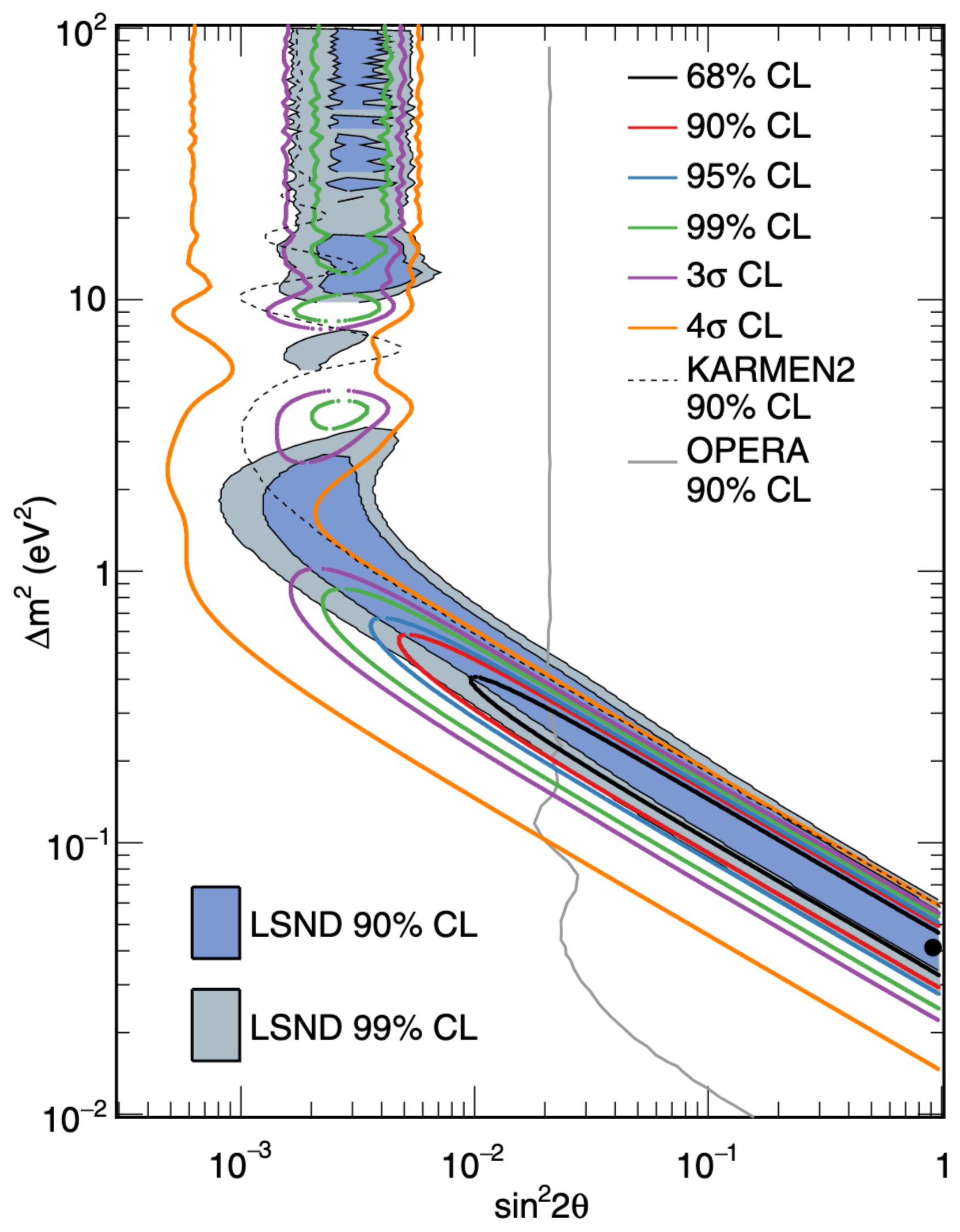}
\end{center}
\caption{
Left: Appearance/prediction vs. L/E shows LSND and MiniBooNE results agree well.  
Right: Allowed regions by LSND and MiniBooNE mostly overlap but OPERA and KARMEN2 exclude some of the allowed regions. Adapted from ~\cite{c:MB}.
}
\label{f:MB}
\end{figure}

MiniBooNE is a 800~ton mineral oil Cherenkov detector which detects neutrinos at a 500~m baseline from pion decay in flight from the Booster neutrino beam (E$_{\nu} = $500~MeV) line at Fermilab. By taking additional 6.38$\times$10$^{20}$ POT $\nu$ data since 2015, so far MiniBooNE has taken total 3$\times$10$^{21}$ POT with almost same amount of $\nu$ and $\overline{\nu}$ for more than 15 years. Using this data MiniBooNE observed excess of $\nu_{\rm{e}}$ and $\overline{\nu}_{\rm{e}}$ events, 460.5 $\pm$ 95.8, corresponding to 4.8$\sigma$, and this result is consistent with LSND excess (see the left plot in Fig.~\ref{f:MB})~\cite{c:MB}. 

The best fit sin$^22\theta$ value for active to sterile neutrino oscillation is close to 1 and this is too big to explain non-observation of sterile neutrinos from OPERA and a combined result of MINOS and MINOS+ that gives best limit on sin$^22\theta_{24}$ for $\Delta m^2 <$ 0.3 eV$^2$~\cite{c:minos+}. 
The right plot in Fig.~\ref{f:MB} shows the allowed (excluded) regions by MiniBooNE and LSND (OPERA and KARMEN2)~\cite{c:MB}. Additionally, the origin of the excess (signal vs. unknown background) events must be identified by MicroBooNE which can distinguish electron-like and gamma-like events.

In addition to the MicroBooNE, the two other detectors in Fermilab SBN program~\cite{c:sbn} using the Booster neutrino beam, ICARUS and ND using the same technology as MicroBooNE, are expected to take data in 2020 or 2021. Therefore, the long time mystery of the LSND result would be expected to be resolved soon.



\subsection{JSNS$^{2}$}

JSNS$^2$ is a 50 ton GdLS (0.2\% Gd) detector in a cylindrical vessel surrounded by gamma catcher region filled with LS surrounded by buffer region filled with mineral oil where 150 10 inch PMTs are mounted on the wall resulting in $\sim$15\% energy resolution at 1~MeV. The outermost region is a muon veto region filled with water. 
JSNS$^2$ is located in the 3$^{\rm{rd}}$ floor of MLF building, JPARC, wehere 3~GeV proton beam is interacted in a mercury target. Unlike MiniBooNE, JSNS$^2$ uses a similar base line (24~m) and the same neutrino source (E$_{\nu} = $30~MeV) from muon decay at rest as LSND. 
JSNS$^2$ is expected to take data in 2020 and with 3 years of data it can exclude LSND result at $\Delta m^2 > 2$ MeV$^2$ region but some phase space is still allowed at $\Delta m^2 < 2$ MeV$^2$. 

\section*{Acknowledgments}
This work was supported by the National Research Foundation of Korea (NRF) grant funded by the Korea Ministry of Science and ICT (MSIT) (No. 2017R1A2B4012757 and IBS-R016-D1-2019-b01).


\end{document}